\documentclass[pdftex,twocolumn,epjc3]{svjour3}          

\RequirePackage{graphicx}
\RequirePackage{mathptmx}      
\journalname{Eur. Phys. J. C}
\usepackage{isotope}
\usepackage{amsmath}
\usepackage{lineno}
\begin{document}
\title{Search for tri-nucleon decays of \isotope[76]{Ge} in GERDA}

\author{
The \mbox{\protect{\sc{Gerda}}} collaboration\thanksref{corrauthor}
\and  \\[4mm]
M.~Agostini\thanksref{UCL} \and
A.~Alexander\thanksref{UCL} \and
G.~Araujo\thanksref{UZH} \and
A.M.~Bakalyarov\thanksref{KU} \and
M.~Balata\thanksref{ALNGS} \and
I.~Barabanov\thanksref{INRM} \and
L.~Baudis\thanksref{UZH} \and
C.~Bauer\thanksref{HD} \and
S.~Belogurov\thanksref{ITEP,INRM,alsoMEPHI} \and
A.~Bettini\thanksref{PDUNI,PDINFN}
% ...
}
\authorrunning{the \textsc{Gerda} collaboration}
\thankstext{corrauthor}{
  \emph{correspondence:}  gerda-eb@mpi-hd.mpg.de}
\thankstext{alsoMEPHI}{\emph{also at:} NRNU MEPhI, Moscow, Russia}
\thankstext{nowDuke}{\emph{present address:} Duke University, Durham, NC USA}
\thankstext{alsoLev}{\emph{also at:} Moscow Inst. of Physics and Technology,
  Russia}
\thankstext{nowIKZ}{\emph{present address:} Leibniz-Institut f{\"u}r
  Kristallz{\"u}chtung, Berlin, Germany}
\thankstext{alsoDubna}{\emph{also at:} Dubna State University, Dubna, Russia}
%
%
%
\institute{
INFN Laboratori Nazionali del Gran Sasso, Assergi, Italy\label{ALNGS}
% ...
\and
Max-Planck-Institut f{\"u}r Kernphysik, Heidelberg, Germany\label{HD} \and
Department of Physics and Astronomy, University College London, London, UK\label{UCL}
% ...
\and
Institute for Nuclear Research of the Russian Academy of Sciences, Moscow, Russia\label{INRM} \and
Institute for Theoretical and Experimental Physics, NRC ``Kurchatov Institute'', Moscow, Russia\label{ITEP} \and
National Research Centre ``Kurchatov Institute'', Moscow, Russia\label{KU}
% ...
\and
Dipartimento di Fisica e Astronomia, Universit{\`a} degli Studi di 
Padova, Padua, Italy\label{PDUNI} \and
INFN  Padova, Padua, Italy\label{PDINFN}
% ...
\and
Physik-Institut, Universit{\"a}t Z{\"u}rich, Z{u}rich, Switzerland\label{UZH}
}

\date{Received: date / Accepted: date}

\maketitle

\begin{abstract}

We search for tri-nucleon decays of \isotope[76]{Ge} in the dataset from the GERmanium Detector Array (GERDA) experiment.
Decays that populate excited levels of the daughter nucleus above the threshold for particle emission lead to disintegration and are not considered. The
ppp-, ppn-, and pnn-decays lead to \isotope[73]{Cu}, \isotope[73]{Zn}, and \isotope[73]{Ga} nuclei, respectively.
These nuclei are unstable and eventually proceed by the beta decay of \isotope[73]{Ga} to \isotope[73]{Ge} (stable).
We search for the \isotope[73]{Ga} decay exploiting the fact that it dominantly populates the 66.7 keV  \isotope[73m]{Ga}
state with half-life of 0.5 s.
The nnn-decays of \isotope[76]{Ge} that  proceed via \isotope[73m]{Ge} are also included in our analysis.
We find no signal candidate and place a limit on the sum of the
decay widths of the inclusive tri-nucleon decays that corresponds to a
lower lifetime limit of 1.2$\times$10$^{26}$ yr  (90\% credible interval). This result improves
previous limits for tri-nucleon decays by one to three orders of magnitude.

\end{abstract}

\section{Introduction}
The Standard Model in its current form appears to conserve baryon number $B$ in all particle interactions. This can be considered as an empirical accidental symmetry. Violation of this symmetry is one of the three Sakharov conditions \cite{Sakharov:1967dj} necessary to explain the observed matter-antimatter asymmetry in the universe.

With a careful choice of the charge assignments for the Standard Model fermions and Higgs boson the Lagrangian is invariant under $Z_6$. Under the chosen charge assignment $Z_6$ is a subgroup of the $U(1)_{2Y-B+3L}$ gauge group and hence any processes must satisfy the condition \cite{babu2003gauged}: \[2\Delta Y - \Delta B +3\Delta L = 0\pmod 6 \] where $Y$ denotes hypercharge and $\Delta Y$ is 0. $L$ denotes the lepton number. It is then apparent that the only valid solutions for this condition require $\Delta B$ be a multiple of 3. As a result $\Delta B=1$ and $\Delta B=2$ processes are suppressed while $\Delta B=3$ processes can occur via dimension 15 operators \cite{babu2003gauged}. Current limits on the proton lifetime are in the order of 10$^{34}$\,{yr} \cite{Workman:2022ynf} which could be attributed to the aforementioned symmetry.

\begin{figure}[ht]
\centering
  \includegraphics[width=.49\textwidth]{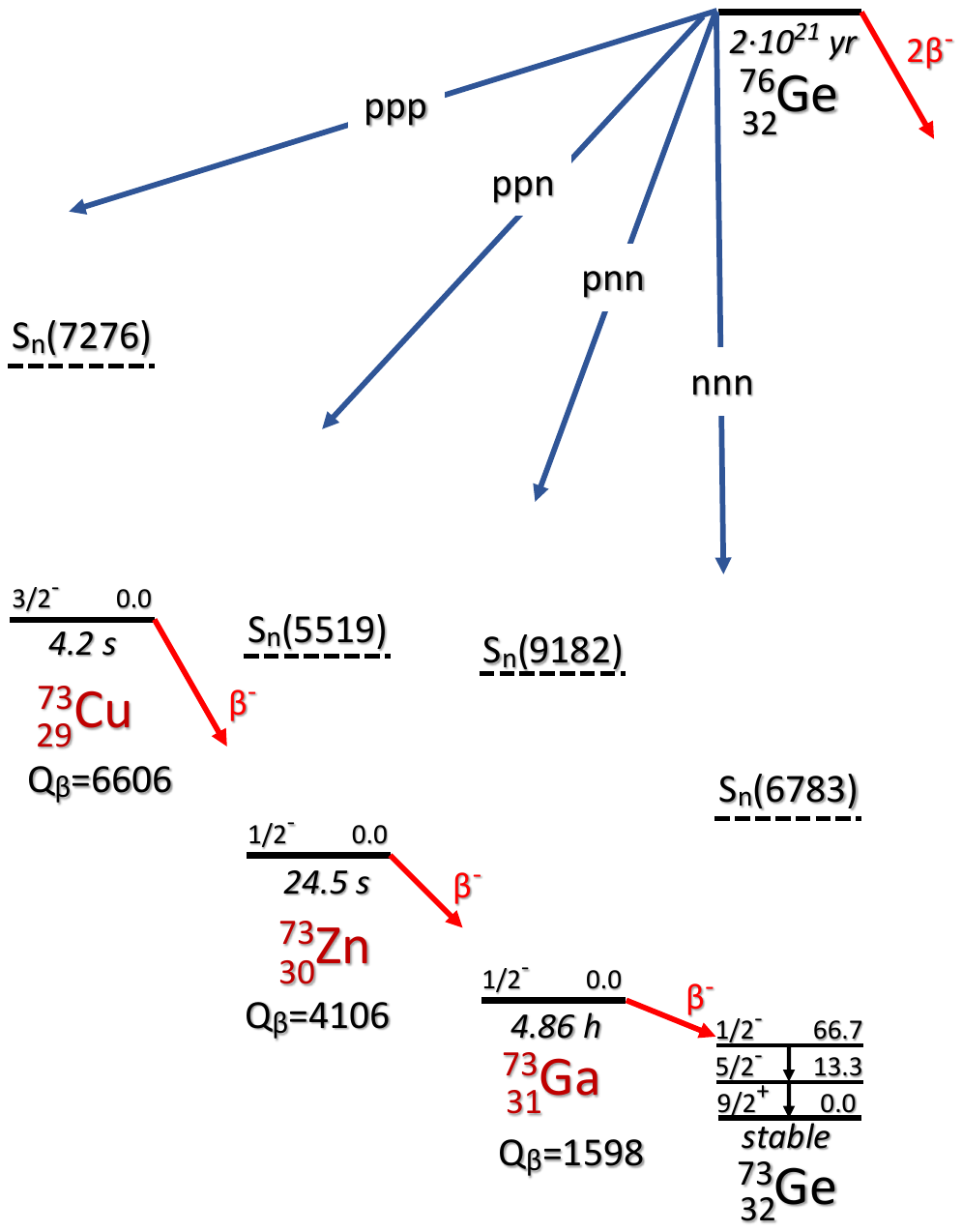}
\caption{Scheme of the potential channels for tri-nucleon decay of 
\isotope[76]{Ge} including the beta decays for the unstable daughter nuclei along 
with their half lives, beta decay Q values and neutron thresholds 
(all energies in keV and not to scale). Also shown are the metastable levels of 
\isotope[73]{Ge} with energies of {66.7}\,keV and {13.3}\,keV and half lives of 
{0.499}\,s and {2.95}\,$\mu$s, respectively. Figure adapted from \cite{osti_1594870}.}
\label{fig:1}       
\end{figure}

The disappearance of three nucleons from a \isotope[76]{Ge} nucleus will spawn A = 73 daughter 
nuclei unless additional nucleons or nuclear clusters are emitted by the daughters; 
hence the total decay width $\Gamma_3^{tot}$  is the sum of two partial decay widths 
$\Gamma^c_3$ and $\Gamma^b_3$ that quantify the population of the continuum and the bound state 
region in the daughter nuclei by tri-nucleon decay. 
Figure~\ref{fig:1} shows all potential tri-nucleon decay channels x, (x = ppp, ppn, pnn, nnn) 
of \isotope[76]{Ge} and for each daughter nucleus the neutron threshold $S_n$ which is 
for \isotope[73]{Cu} and \isotope[73]{Zn} the lowest threshold for particle emission; with 
qualification this holds also for \isotope[73]{Ga} ($S_p=8.843$ MeV, $S_\alpha=6.388$ MeV) 
and  \isotope[73]{Ge} ($S_\alpha=5.305$ MeV) taking the Coulomb barrier for protons 
($\approx$\,6.5 MeV) and $\alpha$ particles ($\approx$\,11 MeV) into account. 
In this study we are concerned with inclusive tri-nucleon decays which populate with partial 
widths $\Gamma^b_x$ just the bound states, i.e. the levels which are stable against particle 
emission, in the daughters, $\Gamma_3^{b} = \sum_x {(\Gamma_x^b)}$, (x\,=\,ppp, ppn, pnn, nnn). 

The decays of the ground states of the three unstable daughters \isotope[73]{Cu}, 
\isotope[73]{Zn}, and \isotope[73]{Ga} proceed all via beta decay: \isotope[73]{Cu} decays 
to \isotope[73]{Zn} with a half life of {4.2}\,s, \isotope[73]{Zn} decays to \isotope[73]{Ga} with 
a half life of {24.5}\,s, and \isotope[73]{Ga} decays to \isotope[73]{Ge} with a half life 
of {4.86\,{h}.
Hence we perform an inclusive search since no assumption is made on the specific type of particles that are produced in the tri-nucleon decay. The only assumption is that the daughter nucleus remains intact.
The tagging of \isotope[73]{Ga} beta decay with the 66.7 keV metastable state of \isotope[73]{Ge}  allows to probe simultaneously the pnn- as well as the ppn- and ppp-channels (see section 3). This tagging includes nnn-decays without 
ionizing particle emission, i.e. 'invisible decays' \cite{heeck2020inclusive}, to the subset of bound states in \isotope[73]{Ge} that decay through the 66.7 keV metastable state.  
Hence we perform in the nnn-channel a semi-inclusive search. The constraint to nnn-decays that are invisible in our detectors ensures the absence of pile-up between the signal from nnn-decay and the tagging event, i.e. a gamma transition in \isotope[73]{Ge}.
For the corresponding partial decay width holds  $\Gamma_{nnn}^{b'} < \Gamma_{nnn}^{b}$. Our measurement will constitute thus the limit for the sum of partial decay widths
$\Gamma^b_3 = \Gamma_{ppp}^b + \Gamma_{ppn}^b + \Gamma_{pnn}^b +\Gamma_{nnn}^{b'}$ of the tri-nucleon process with a partial lifetime limit $\tau_b = 1/\Gamma^b_3$.


\section{The GERDA experiment}
The GERmanium Detector Array (GERDA) experiment \cite{ackermann2013gerda,agostini2018upgrade} was located at the Laboratori Nazionali del Gran Sasso (LNGS) of INFN under the Gran Sasso mountain, Italy. The overhead rock provides shielding from atmospheric muons with a mean muon flux 
of 3.5x10$^{-4}$\,s$^{-1}$m$^{-2}$ \cite{agostini2016flux}. The experiment employed High Purity Germanium (HPGe) detectors in a liquid argon (LAr) cryostat \cite{knopfle2022design} housed within a tank of ultra-pure water instrumented with photomultiplier tubes (PMTs) to tag the Cherenkov light from incident muons. GERDA employed several types of HPGe detectors with different geometries: semi-coaxial and BEGe (Broad Energy Germanium) \cite{agostini2017background} and inverted coaxial \cite{agostini2021characterization}. Above the cryostat was a lock system accessed via a clean room which isolated the LAr in the cryostat from the lab atmosphere and allowed for the insertion and removal of strings of detectors. Figure~\ref{fig:2} shows a cross section of the installation with these key features.
\begin{figure}[ht]
\centering
 \includegraphics[width=.4\textwidth]{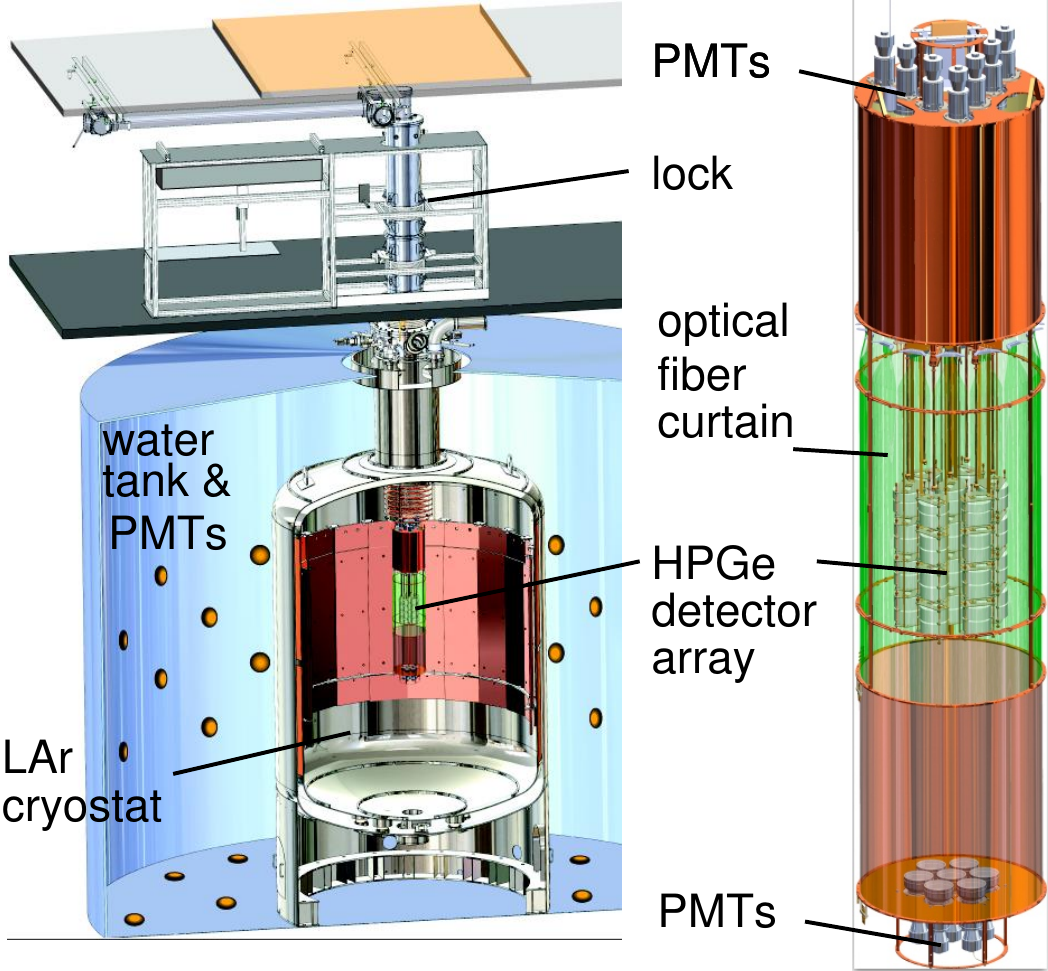}
\caption{Cross sections of the GERDA experimental apparatus and an enlarged view of the central part (right), the germanium detector array enclosed by the LAr veto system \cite{agostini2018upgrade}.}
\label{fig:2}       
\end{figure}

GERDA’s primary purpose was to search for a signature of neutrinoless double beta 
($0\nu \beta \beta$) decay to probe the Majorana nature of the neutrino \cite{agostini2020final}. 
A good candidate source nucleus for $0\nu \beta \beta$ decay must not undergo single beta decay. 
This condition is fulfilled by \isotope[76]{Ge} for which the process is energetically forbidden. 
For this search then, GERDA's HPGe detectors are enriched in \isotope[76]{Ge} to about 88\% and 
operated directly in LAr. The LAr has a dual purpose, to cool the detectors to cryogenic temperatures
as well as provide a veto system from the scintillation light of processes depositing energy in 
the LAr. Scintillation light is detected by PMTs or silicon photomultipliers (SiPMs) coupled to 
wavelength shifting fiber shrouds. The LAr is also used to shield against external background 
contributions from \isotope[238]{U} and \isotope[232]{Th} decay chains. In both $0\nu \beta \beta$ 
and tri-nucleon decay searches \isotope[76]{Ge} is both the detector and source material allowing 
for excellent detection efficiency. Further, the exceptional energy resolution obtainable by 
HPGe detectors is well established in the literature 
\cite{agostini2017background,agostini2021characterization,agostini2021calibration}.


\section{Tagging $\mathbf{^{73}Ga}$ beta decays via the $\mathbf{^{73m}Ge}$ decay}  
\isotope[73]{Ga} is an unstable isotope that beta decays with a half life of {4.86}\,{h} to excited states of the stable isotope \isotope[73]{Ge}. Figure~\ref{fig:3} shows the level scheme for \isotope[73]{Ge} populated by this process. Importantly, the initial beta decay does not populate the \isotope[73]{Ge} ground state due to a large nuclear spin difference compared to the \isotope[73]{Ga} ground state ($9/2^+$ and $3/2^-$ respectively). 5.9\% of decays will directly populate the $1/2^-$ metastable state at {66.7}\,keV, all other decays will populate higher energy levels which will decay by gamma emission. Virtually all cascades will transition to the metastable state. Altogether 98.2\% of \isotope[73]{Ga} decays will promptly reach the metastable state which functions as a bottleneck in the decay to the \isotope[73]{Ge} ground state. When an event is triggered in the detector the flash ADC is readout for a trace window of {160}\,$\mu$s, the waveform's leading edge is centred in the trace window at {80}\,$\mu$s.
\begin{figure}[hb]
\centering
\includegraphics[width=.5\textwidth]{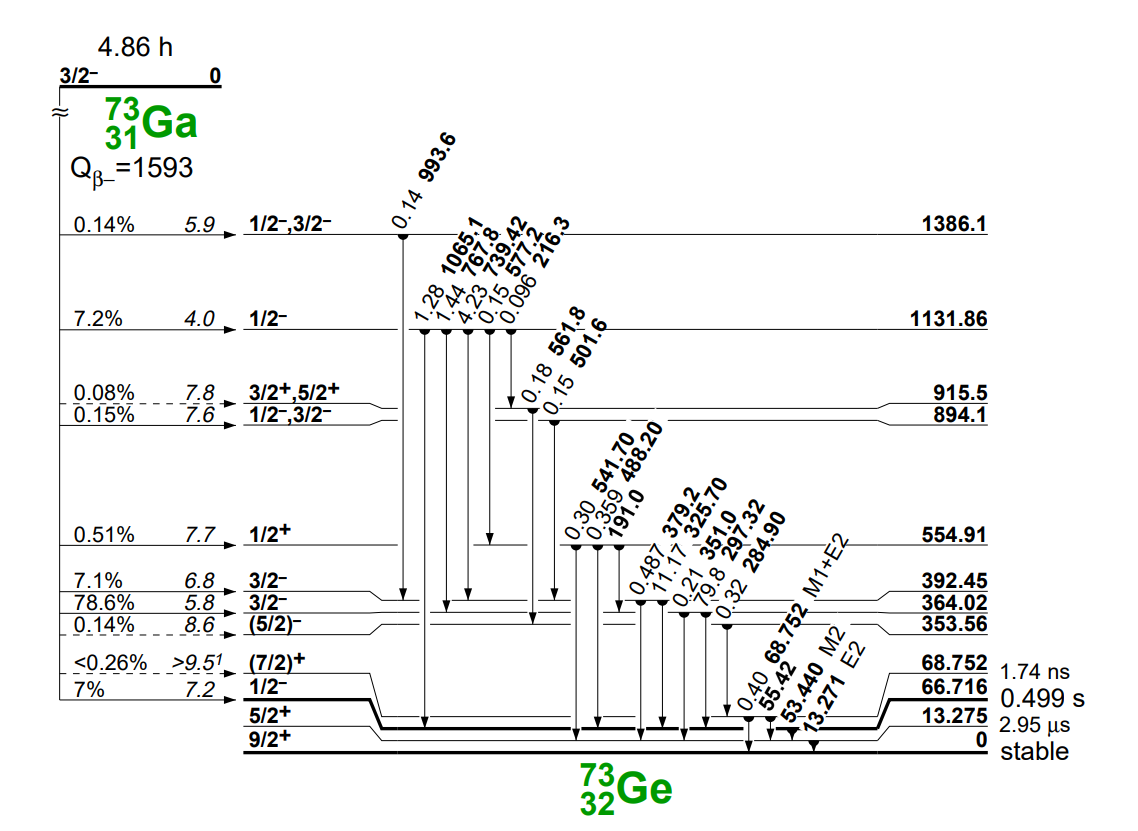}
\caption{Energy levels of \isotope[73]{Ge} populated by \isotope[73]{Ga} beta 
decay \cite{firestone1998table};
see \cite{osti_1594870} for an update of some branching ratios, 
e.g. 5,9\% to the {66.7}\,keV level. The ground state of \isotope[73]{Ge} is never 
directly populated.}
\label{fig:3}       
\end{figure}

The initial beta decay and subsequent gamma cascade will trigger an initial event 
in the detector corresponding to their summed energies. In the analysis we call 
this sum $E_1$. Since the metastable state has a half life of {0.499}\,s its decay 
will constitute  a separate event trigger with energy {66.7}\,keV, which we will 
refer to as $E_2$, in practically all decays. 
The energy of the metastable state is such that only a subset of the GERDA Phase II 
data in which the trigger threshold was lowered to around {20}\,keV could be 
considered.

The metastable state decays via a two step cascade. Firstly the {66.7}\,keV state decays to the {13.3}\,keV state emitting {53.4}\,keV. This {13.3}\,keV state is also metastable but with a much shorter half life of {2.95}\,$\mu$s, small compared to the recorded trace window of {160}\,$\mu$s, and will decay shortly afterwards by emitting {13.3}\,keV. The Figure~\ref{fig:4} shows a typical waveform from the decay of the 66.7 keV state. 
Whereas a waveform from background events normally exhibits a single leading edge to a maximum amplitude, the waveform of the 66.7 keV metastable state will have two leading edges. The metastable state was previously used in analyses of muon activity in MAJORANA also making use of the unique shape \cite{arnquist2022signatures}.
\begin{figure}[htb]
\centering
\includegraphics{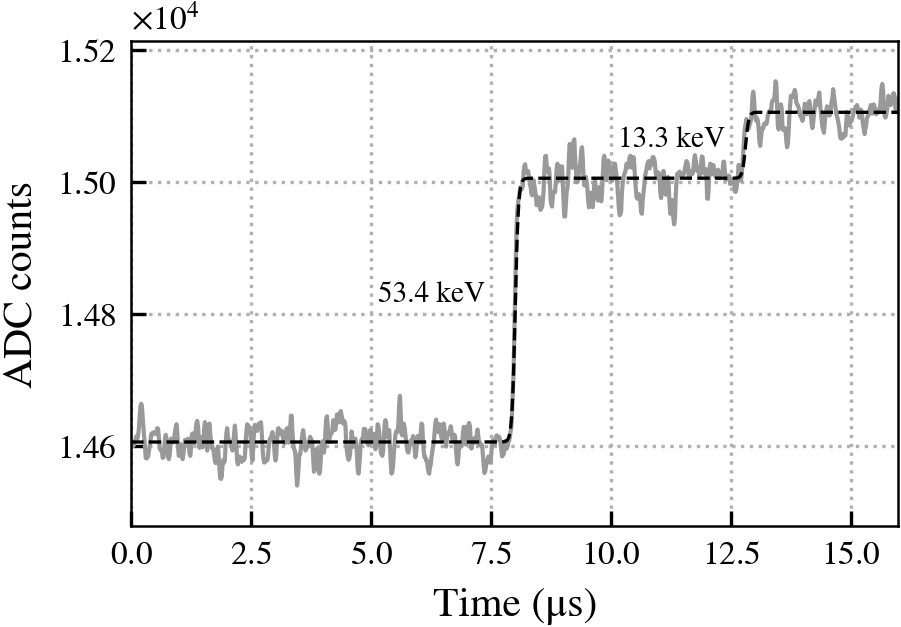}
\caption{Simulated waveform of the decay of the {66.7}{keV } metastable state in \isotope[73]{Ge} to the ground state via the intermediate {13.3}\,keV state. Details of the response of the amplifier to the signal are omitted.}
\label{fig:4}       
\end{figure}

Our search procedure for the \isotope[73]{Ga} decay is then to consider by delayed coincidences any pair of events with energies $E_1$ and $E_2$ recorded in a single detector within {2.5}\,s of each other (5 half lives of the metastable state) where the first event has the energy $E_1$  below the Q value ({1598}\,keV) of the \isotope[73]{Ga} beta decay. 
  
 
\section{Monte Carlo simulation of $^{73}$Ga decays }
The responses to $^{73}$Ga decays originating from inside the HPGe detectors were simulated using the MaGe software package \cite{boswell2011mage} based on Geant4 \cite{agostinelli2003geant4}. MaGe does not provide waveforms of events but instead records energy depositions with position and timing information in defined sensitive regions of the detectors. The timing information was used to group energy depositions within {80}\,$\mu$s windows (half of a trace length for data waveforms) which separated the simulated $E_1$ and $E_2$ events. Following the clustering a smearing is applied to the energies to simulate the  energy resolution of the detectors. The parameters for this smearing are obtained from the energy resolution curves for each detector type \cite{agostini2021calibration}. For the Monte Carlo simulation  $10^7$ primaries were simulated.

Figure~\ref{fig:5} shows the spectrum for the first energy depositions in each detector after time clustering with the expected continuous energy distribution. Also modelled in the Monte Carlo simulation is the detector dead layer, a region of no  or partial charge collection in the outer detector layers \cite{agostini2019characterization}. The dead layer is modelled here as a hard transition at a depth close to {1}{mm}. We observe that 40\% of these $E_1$ energy depositions also deposited energy in the LAr, thus applying the LAr veto to these energies in the search would lead to a significant loss of signal. On the other hand, in the decay of the metastable state rarely energy is deposited in the LAr and the LAr veto can be applied.
Figure~\ref{fig:6} shows the spectrum for the second energy depositions corresponding to $E_2$ after time clustering. It is dominated by the peak at {66.7}\,keV which contains about 99\% of all events. Additional peaks at {13.3}\,keV and {53.4}\,keV and the continuum below {66.7}\,keV are due to one of the transitions in the two step decay depositing some energy in the dead layer or escaping to the LAr.
\begin{figure}[ht]
\centering
\includegraphics{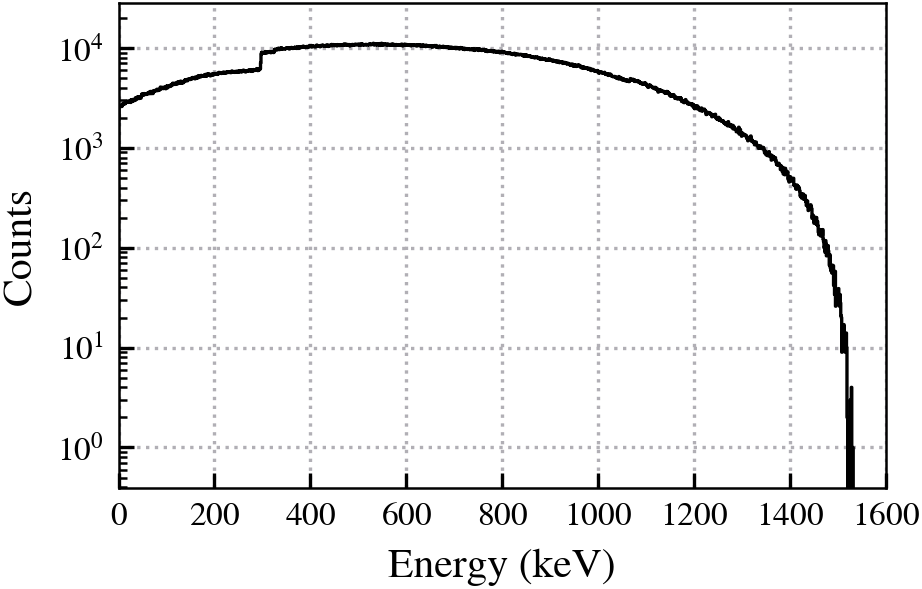}
\caption{Monte Carlo energy spectrum corresponding to $E_1$, the energy of the 
\isotope[73]{Ga} decay to \isotope[73]{Ge} and the subsequent gamma transition to its 
metastable state. The step at about {300}\,keV can be explained by the level scheme 
of \isotope[73]{Ge}: 6\% of \isotope[73]{Ga} decays directly populate the metastable 
state. 78.6\% of decays will populate a state at {364}\,keV which can transition to the 
metastable state releasing ~{297}\,keV which accounts for the step.} 
\label{fig:5}       
\end{figure}

The energy difference $|E_1 - E_2|$ between the prompt and delayed event can be used to effectively discriminate signal and background. The primary accidental background is from beta decays of \isotope[39]{Ar} with a Q value of 565 keV. In these events the difference between the prompt and delayed energy will on average be small compared to the difference for signal events as the energies come from the same distribution. For signal events the prompt energy is on average much greater than the delayed energy, hence the energy difference is greater than in background.  
\begin{figure}[t]
\centering
\includegraphics{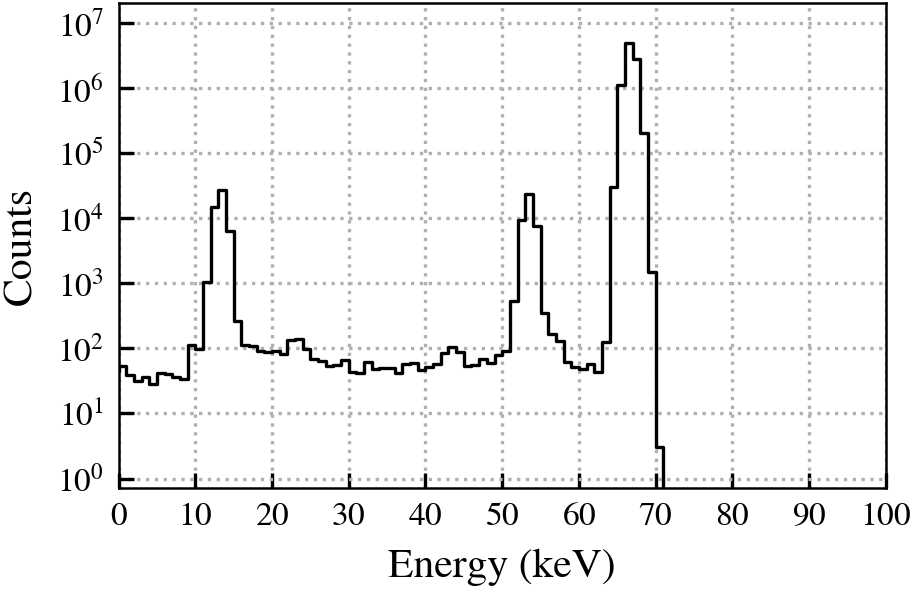}
\caption{Monte Carlo energy spectrum corresponding to $E_2$, the energy released in the decay 
of the {66.7}\,keV metastable state of \isotope[73]{Ge}. Approximately 99\% of entries are 
contained within the peak at {66.7}\,keV.}
\label{fig:6}       
\end{figure}

The cumulative distribution function (CDF) for the spectrum of the difference between $E_{1}$ and $E_2$ was obtained from the Monte Carlo data. The same CDF was obtained for the background using GERDA data after anti-coincidence and quality cuts to select physical events occurring within a single detector. We assume that the ratio of signal to background events in the GERDA data is negligible and the data can be treated as purely background for the purpose of both energy and risetime distributions (see section 5 for further details). These CDFs were used to optimise an energy cut using the product of signal efficiency and background rejection as a performance metric. Both CDFs and the metric are shown in figure~\ref{fig:7}.
\begin{figure}[b]
\centering
\vskip-0.3
truecm
\includegraphics{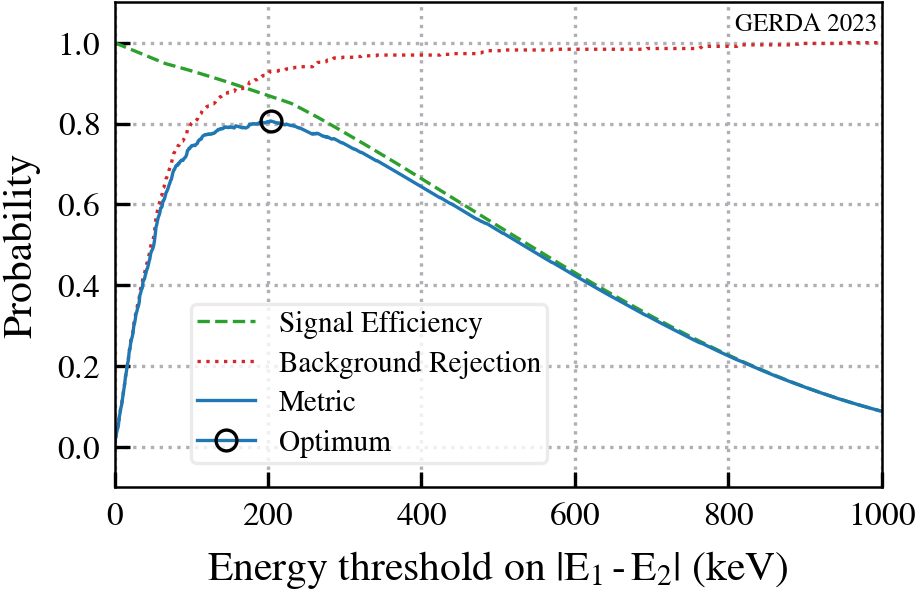}
\caption{Energy cut optimisation plot showing the background rejection, the signal survival fractions and the performance metric as a function of the energy threshold $|$$E_1 - E_2$$|$. The signal efficiency for the optimum is 87\% with 93\% background rejection at an energy difference threshold of {204}\,keV. The metric curve is the product of the signal and background curves.}
\label{fig:7}       
\end{figure}

\section{Risetime cuts}
We reconstruct from the GERDA data both $\tau_{60}$ and $\tau_{90}$ for waveforms. These are the times taken for a waveform to go from 5\% to 60\% and from 10\% to 90\% of its maximum amplitude respectively. For the two step $E_2 = 66.7$\,keV event the 60\% threshold is {40}\,keV and occurs during the first leading edge. The 90\% threshold is {60}\,keV and occurs during the second leading edge. As a result $\tau_{90}$ is proportional to the survival time of the {13.3}\,keV intermediate state. 
Typical $\tau_{90}$ values are on the order of a few {100}\,ns for background events, while the {13.3}\,keV state's half life is an order of magnitude larger. Signal events generally have significantly longer $\tau_{90}$ values than typical events in the data. $\tau_{60}$ on the other hand has no dependence on the survival time of the {13.3}\,keV state and expected $\tau_{60}$ values for signal and background events are comparable. Using the risetime information we optimised a threshold cut on the composite variable $\tau_{90} - \tau_{60}$, which is large for signal waveforms and small for background and hence a clear separation is expected between the two distributions. The parameter also implicitly applies a threshold on $\tau_{90}$ which is larger for signal events. This risetime optimisation was performed individually for each detector type employed in GERDA with figure~\ref{fig:8} showing an optimisation plot for the BEGe detectors.
\begin{figure}[ht]
\centering
\vskip-0.35truecm
\includegraphics{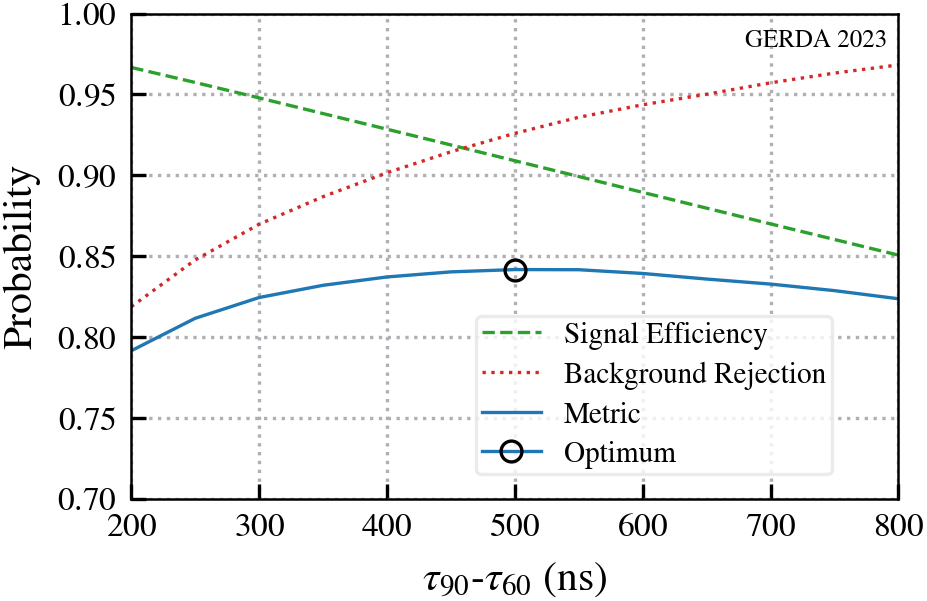}
\caption{Risetime optimisation plot for BEGe detectors. The signal efficiency for the optimum is 91\% with 93\% background rejection at a time difference threshold of {500}\,ns. The other detector types exhibit similar performance.}
\label{fig:8}       
\end{figure}

\section{Results and discussion}
Figure~\ref{fig:9} shows on top the energy distribution of the GERDA events between 20 keV  and 1600 keV before analysis cuts. The primary background for this search comes from the beta decay of \isotope[39]{Ar}. The mean overall event rate in a detector is approximately 1 event every 10 minutes. 44 detectors were considered in this analysis. 
We applied the search criteria discussed in section 3 along with the energy and risetime cuts as well as with the LAr veto for the 2nd event of the delayed coincidences. A summary of the cuts with efficiencies and the detector-type independent efficiencies are shown in table~\ref{tab:3}. We obtained a histogram of the surviving  event energies $E_2$ shown in figure~\ref{fig:9} bottom.
\begin{table}[t]
\centering
\caption{Summary of energy $E_{1,2}$, rise time $\tau_{60/90}$ and LAr veto cuts in the delayed coincidence between events 1 and 2 within the time window $T_2 - T_1$. Corresponding efficiencies are denoted by $\epsilon$. For the nnn-decay search the $E_1$ cut has been relaxed to 6.8 MeV. The region of interest for $E_2$ is 40 - 72 keV. }
\label{tab:3}       
\begin{tabular}{lll}
\hline\noalign{\smallskip}
Cut & value~~~~~ & $\epsilon$ \\
\noalign{\smallskip}\hline\noalign{\smallskip}
  $E_1, E_2$ trigger threshold & $\sim$20 keV & 1 \\
  $E_1$ & $<1600$ keV & 1 \\
  1st event: LAr veto & no & - \\
2nd event: LAr veto & yes & 0.975 \\ 
 $|$$E_1 - E_2$$|$&  $>204$ keV  &  0.870 \\
 $T_2 - T_1$ & $<2.5$ s &  0.969 \\
 2nd event: $\tau_{90}$ - $\tau_{60}$ & $>500\,(400^*)$ ns  & see $\epsilon_\tau$ in table 2 \\
\noalign{\smallskip}\hline
* for coaxial detectors
\end{tabular}
\end{table} 
\begin{figure}[b]
\centering
\includegraphics{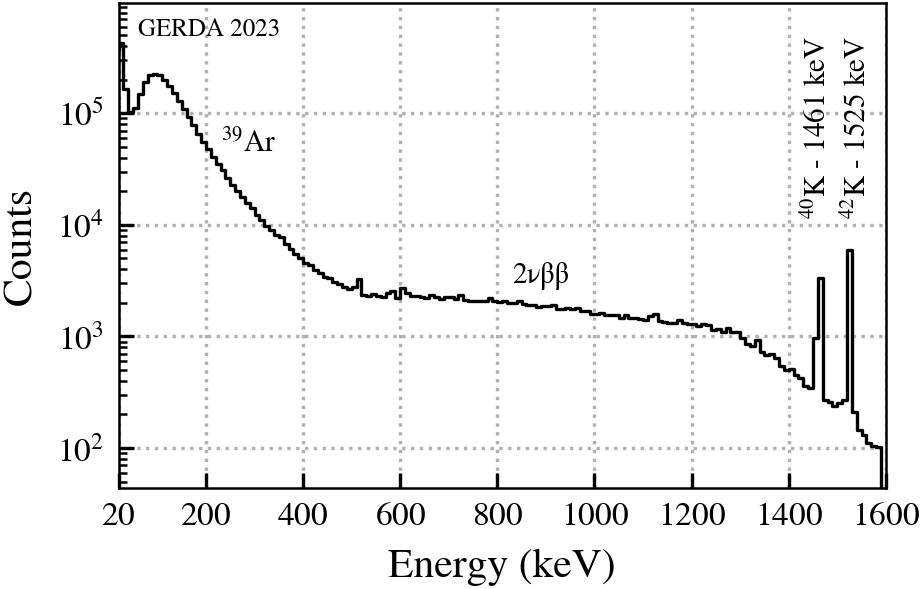}
\includegraphics{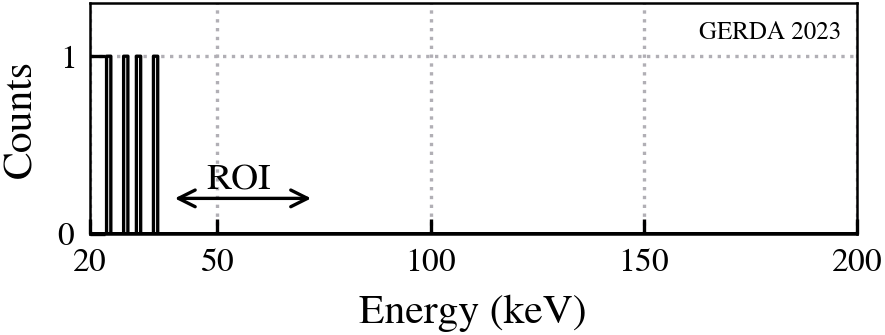}
\caption{GERDA energy spectrum from {20}\,keV to {1600}\,keV before any analysis cuts. 
The contributions from  \isotope[39]{Ar}, two neutrino double beta (2$\nu\beta\beta$) decay 
and  \isotope[40,42]{K} are indicated. Bottom: Surviving $E_{2}$ events after applying our 
search procedure with energy, rise time and LAr veto cuts. 
The region of interest (ROI) is indicated. The expected number of accidental events up to 
1600 keV is about 2.}
\label{fig:9} 
\end{figure}
We considered a search region of 40\,keV to 72\,keV for the 66.7 keV signal. The upper limit of {72}\,keV is due to the energy resolution of HPGe detectors (energy resolution at 67~keV $\leq 5$~keV at full width half maximum). The lower bound arises from the possibility that some energy is lost in the dead layer or the reconstructed energy is lower since the energy integration time of around 1~$\mu$s is lower than the half-life of the 13.3~keV state. The lower value of {40}\,keV is  well above the trigger threshold. The limited energy integration time does not influence the risetime reconstruction. There are 4 events surviving our cuts all of which are below {40}\,keV. No candidates survive in the region of interest. This observation holds even if we abandon the $E_1 < 1600$ keV cut such that any gamma cascade from the \isotope[73]{Ge} bound state region up to 6.8 MeV is accepted. Hence we find no evidence for tri-nucleon decays of \isotope[76]{Ge} to the bound states of \isotope[73]{Cu}, \isotope[73]{Zn}, \isotope[73]{Ga}, and for invisible nnn-decays to \isotope[73]{Ge}.

From the exclusion of the \isotope[73]{Ga} decay in our dataset we set a lower limit on the partial mean lifetime of the tri-nucleon process at 90\% credibility.
The exposures for each detector type and the combined analysis exposure are shown in table~\ref{tab:1} along with associated analysis efficiencies. The total exposure is {61.89}\,{kg\,yr} with a combined analysis efficiency $\epsilon_{tot}$ of 0.554.
\begin{table}[h]
\centering
\caption{Summary of exposures by detector type with their respective analysis efficiencies. The last row shows the exposure-weighted sums of the above efficiencies. 
$\mathcal{E}$ denotes the exposure in {kg\,yr}, 
$\epsilon_{\tau}$ denotes the risetime cut efficiency, 
$\epsilon_{v}$ denotes the active volume fraction, and $\epsilon_{e}$ denotes the enrichment fraction. 
$\epsilon_{tot}$ is defined as the product of all analysis efficiencies.
The additional efficiency terms that do not depend on the detector type and 
contribute to $\epsilon_{tot}$ are shown in table~\ref{tab:3} and include the fraction 
of beta decays populating the metastable 66.7 keV state (0.982). }
\label{tab:1}       
\begin{tabular}{llllll}
\hline\noalign{\smallskip}
Type & $\mathcal{E}$ {(kg\,yr)} & $\epsilon_\tau$ & $\epsilon_v$ & $\epsilon_e$ & $\epsilon_{tot}$  \\
\noalign{\smallskip}\hline\noalign{\smallskip}
Inverted Coax & 8.30  & 0.901 & 0.926 & 0.877 & 0.591 \\ 
BEGe             & 30.96 & 0.909 & 0.886 & 0.877 & 0.570 \\ 
Coax          & 20.67 & 0.933 & 0.867 & 0.864 & 0.564 \\ 
Natural Coax  & 1.96  & 0.927 & 0.854 & 0.078 & 0.050 \\ 
Combined         & 61.89 & 0.917 & 0.884 & 0.847 & 0.554 \\ 
\noalign{\smallskip}\hline
\end{tabular}
\end{table} 
Disregarding contributions from nnn-decays a conservative partial lifetime limit 
for the tri-nucleon processes 
x = ppp, ppn, and pnn  was then calculated with the following formula
\begin{align}
\mathrm{\tau_{b}} \ge \frac{1}{S} \frac{N_a}{m_{Ge}} 
                       \sum\limits_{i}^{}\mathcal{E}_i{\epsilon_{tot_{i}}}
\end{align}
where $\mathrm{\tau_{b}}$ denotes the partial lifetime for tri-nucleon decays to the bound states of the daughter nuclei. $\mathcal{E}_i$ denotes the exposure for a particular detector type. $\epsilon_{tot_{i}}$ denotes the total analysis efficiency for a detector type. $N_a$ is Avogadro's constant, and $m_{Ge}$ is the molar mass of the enriched germanium in the detectors. 
$S$ denotes the lower signal limit at 90\% CI for no observed background or signal and has a value 
of 2.3 counts in a Bayesian analysis. An exposure of {61.89}\,kg\,yr corresponds to a lower 
limit on $\mathrm{\tau_{b}}$ of 1.2x10$^{26}$\,{yr} on the aforementioned tri-nucleon decay channels of \isotope[76]{Ge}. The main systematic uncertainty of this analysis arises from the active volume and enrichment fractions of the detectors (4\%) which do not significantly contribute to the stated limit and can be neglected.

Above analysis can also be applied to the invisible nnn-decays to bound states of \isotope[73]{Ge}. If $k$ denotes the fraction with which these states decay via the 66.7 keV metastable state, the lower limit on the partial lifetime of this process is estimated to be $k\cdot 10^{26}$ yr at 90\% credibility . Hence a fraction $k$ as low as $10^{-3}$ would still constrain the partial mean lifetime of the considered invisible nnn-decays to $10^{23}$ yr. 
 No estimates of the value of k are available. The main challenge in calculating this value is the poorly known level scheme of \isotope[73]{Ge} up to the neutron threshold and the unknown reaction mechanism of nnn-decay of \isotope[76]{Ge}. 

Table~\ref{tab:2} compiles our results together with a summary of current tri-nucleon decay limits. 
We quote for each inclusive \isotope[76]{Ge} decay channel x (x = ppp, ppn, and pnn) the lifetime $\tau_b$[x] corresponding to the summed decay width $\Gamma^b_3 = \sum_x{(\Gamma^b_x)}$; this is conservative since $\Gamma^b_x \leq \Gamma^b_3$ and the corresponding lifetime is the inverse of the respective decay width.
\begin{table}[b]
\centering
\caption{Present results and overview of lower limits of partial lifetimes $\tau_b$[x] for 
indicated decay channel x (x = ppp, ppn, pnn, and nnn)
from previous searches for tri-nucleon decays. Extension \lq+ X\rq \ marks inclusive decay modes. $k$ denotes the fraction of invisible nnn-decays to bound states of \isotope[73]{Ge} that decay via the metastable 66.7 keV state. Note that MAJORANA's results have been converted from the quoted half life limits to mean lifetime limits.}
\label{tab:2}       
\begin{tabular}{lll}
\hline\noalign{\smallskip}
Experiment & decay & $\mathrm{\tau_{b}}$[x] (yr)  \\
\noalign{\smallskip}\hline\noalign{\smallskip}
 GERDA & \isotope[76]{Ge} $\xrightarrow{ppp}$ \isotope[73]{Cu} + X & $1.20\times 10^{26}$ \\
        & \isotope[76]{Ge} $\xrightarrow{ppn}$ \isotope[73]{Zn} + X & $1.20\times 10^{26}$ \\
        & \isotope[76]{Ge} $\xrightarrow{pnn}$ \isotope[73]{Ga} + X & $1.20\times 10^{26}$ \\
        & \isotope[76]{Ge} $\xrightarrow{nnn}$ \isotope[73]{Ge} + X$_{invisble}$& $k \times 10^{26}$ \\
MAJORANA \cite{alvis2019search} & \isotope[76]{Ge} $\xrightarrow{ppp}$ \isotope[73]{Cu} + X & $1.08\times 10^{25}$ \\ 
                   & \isotope[76]{Ge}  $\xrightarrow{ppp}$ \isotope[73]{Cu}\,e$^+\pi^+\pi^+$ & $6.78\times 10^{25}$  \\
                   & \isotope[76]{Ge} $\xrightarrow{ppn}$ \isotope[73]{Zn}\,e$^+\pi$ $^+$ & $7.03\times 10^{25}$ \\ 
 EXO-200 \cite{Xe136}         & \isotope[136]{Xe} $\xrightarrow{ppp}$ \isotope[133]{Sb} + X & $3.3\times 10^{23}$  \\
                              & \isotope[136]{Xe} $\xrightarrow{ppn}$ \isotope[133]{Te} + X & $1.9\times 10^{23}$  \\ 
Hazama et al. \cite{hazama1994limits} & \isotope[127]{I} $\xrightarrow{nnn}$ \isotope[124]{I} + X & $1.8\times 10^{23}$ \\
\hline
\end{tabular}
\end{table}
 Previous limits for \isotope[76]{Ge} were set by the MAJORANA collaboration for the inclusive copper channel and for the
 exclusive copper and zinc channels assuming the quoted decay channels to be the dominant ones without identifying the particular emitted particles \cite{alvis2019search}. Our result improves on the limits for these channels as well as setting the first limits on both the inclusive gallium and zinc channels. In addition, the limits from our analysis have no model dependence concerning the decay channel.
 Inclusive ppp- and ppn-decay studies have also been performed with \isotope[136]{Xe} \cite{Xe136} yielding limits in the order of 10$^{23}$\,{yr}. Inclusive nnn-decays have been searched for with \isotope[127]{I} \cite{hazama1994limits}. The deduced limit in the order of 10$^{23}$\,{yr} is, however, not a nuclear lifetime but takes shell model combinations of baryons within the nucleus into account.
 Current limits for proton \cite{takenaka2020search} and di-nucleon decays \cite{araki2006search} are many orders of magnitude larger providing good motivation to investigate multi-nucleon decays. 
Our results represent the most stringent limits on inclusive tri-nucleon decays to date by utilising the unique properties of the gamma cascade of the de-excitation of the metastable state in \isotope[73]{Ge}. LEGEND (Large Enriched Germanium Experiment for Neutrinoless $\beta\beta$ Decay), GERDA's successor, has started to collect data and will eventually offer a much greater exposure for future tri-nucleon decay searches with \isotope[76]{Ge}. The quantitative study of the nnn-decay to \isotope[73]{Ge} remains a challenge. Moreover, muon induced spallation could also create \isotope[73]{Ga} within HPGe detectors with an identical signature to tri-nucleon decay. Future analyses with higher exposure datasets may then contain such cosmogenic \isotope[73]{Ga} decays, potentially hampering the stringency of limits that can be obtained. In the case of no signal candidates as in our dataset this problem does not manifest.

\begin{acknowledgements}
The \textsc{Gerda} experiment is supported financially by
the German Federal Ministry for Education and Research (BMBF),
the German Research Foundation (DFG),
the Italian Istituto Nazionale di Fisica Nucleare (INFN),
the Max Planck Society (MPG),
the Polish National Science Centre (NCN),
the Foundation for Polish Science (TEAM/2016-2/17),
the Russian Foundation for Basic Research,
and the Swiss National Science Foundation (SNF).
This project has received funding/support from the European Union's
\textsc{Horizon 2020} research and innovation programme under
the Marie Sklodowska-Curie grant agreements No 690575 and No 674896.
This work was supported by the Science and Technology Facilities Council, part
of the U.K. Research and Innovation (Grant No. ST/T004169/1).
The institutions acknowledge also internal financial support.
 
The \textsc{Gerda} collaboration thanks the directors and the staff of
the LNGS for their continuous strong support of the \textsc{Gerda} experiment.
\end{acknowledgements}


\end{document}